# CESSATION OF VOLCANISM ON EARTH-POSSIBILITIES IN NEAR GEOLOGICAL FUTURE

Varnana.M.Kumar[1], T.E.Girish[1], BijuLonghinos[1], Thara.N.Satyan[2] and AnjanaA.V.Panicker[3]

1 Department of Physics, University College, Trivandrum 695034, INDIA

2 Department of Geology, College of Engineering, Trivandrum 695016, INDIA

3 Department of Geology, University College, Trivandrum 695034, INDIA

The number of active volcanoes and its latitudinal extent is likely to be related to the magnitude of internal heat in rocky planets. A critical value of internal heat may require in these planets to sustain volcanism and the decline of volcanic activity since their formation of these planets is inferred to be governed by radioactive decay laws. We find that major volcanic activity in Mars, Moon, Mercury and Venus has ceased when their respective surface heat flux values are within ±10 % of the current surface heat flux value of Earth ($0.093$ W/m$^2$). The reduction in spatial extent of recent volcanism in Venus compared to the geological past is inferred to be part of significant reduction in volcanic activity in this twin planet of Earth. We suggest that the volcanic activity in Earth is also declining significantly since the period of mass extinction of dinosaurs 65 million years ago. It may cease completely within a time span between 19-65 million years from now with possible Implications in Earth's interior, climate and biosphere.

**Key words:** volcanic activity, rocky planets, Earth, cessation, inner solar system, internal heat, mass extinction

# 1 Introduction

Volcanic activity in rocky planetary bodies is maintained by planetary internal heat which in turn can be generated by three different mechanisms i) Primordial heat related to the formation of the planets ii.Radiogenic heat generation in the crust and mantle due to presence of radioactive elements iii.Tidal heating (Pater et al.,2013; Jackson et al.,2008). For the inner solar system planetary bodies generally the first two mechanisms are important. Earth is the only planet for which we have direct observations of internal heat parameters (Davies and Davies, 2010).The surface heat flow/surface heat flux in Earth is currently estimated from the spatially resolved measurements covering most parts of the Earth. Surface heat flow in a rocky planet is generally inhomogeneous along its surface and total surface heat flux defined for the same is a more meaningful physical quantity. Total radiogenic heat generated in the Earth is currently best determined from geoneutrino observations (Korenaga, 2011). For other planetary objects in the inner solar system the internal heat parameters are inferred from surface heat flows using remote sensing observations.

Earth is only rocky planet in the solar system to have both volcanic activity as well as plate tectonic activity at present. Major volcanism in other rocky planetary objects in the solar system has ceased in the geological past. This situation naturally poses some questions viz. (i) Is there any critical value of internal heat required in a rocky planetary object to sustain volcanism (ii) How volcanic activity in rocky planetary bodies like the Earth's Moon has ceased in the geological past ? (iii) Is there any possibility of cessation of major volcanic activity of Earth in the near geological future?

In this paper we have addressed the above questions in detail. We have first inferred internal heat parameters of rocky planetary bodies in the inner solar system from best available observations and models. Using these data we have estimated the time evolution of surface heat flux of Earth, Venus, Mercury, Mars and Moon using radioactive decay laws (Schubert and Stevenson 1980) .These results are compared with volcanic activity history over geological time scales in the above planetary bodies. It is found that major volcanic activity in Mars,Moon,Mercury and Venus has ceased when their respective surface heat flux values are close to the current surface heat flux in Earth ( 0.093 W/m$^2$) with a difference of below 10 %.This may imply that volcanic activity in Earth has started declining already. It is possible that this decline is possibly started 65 million years when mass extinction of dinosaurs happened due to a prolonged and major volcanic eruptions. It is inferred that volcanic activity in earth may cease completely within a time span of about 260 million years from now with possible implications in our climate and biosphere.

## 2 Analysis and Results

### 2.1 History of volcanism inferred for the inner solar system rocky planetary objects

In Table 1, we have given important characteristics of volcanic activity inferred /observed in the geological time scale for the inner system rocky planetary objects: Earth,Venus,Mars,Mercury and Moon. The number and spatial extend of volcanoes; the inferred period of cessation of major volcanic activity in these planetary objects are given in this Table. The volcanism in Earth, Venus is found to extend up to the Polar Regions. Further the number of volcanoes erupted till date in these planets significantly outnumbers the same in Mars, Moon and Mercury. Larger internal heat sustained in Earth and Venus is likely to be the reason for intense and widespread volcanism inferred or observed in this planets.Eventhough major volcanism in Mars and Venus has possibly ceased in the geological past .Minor volcanic activity is inferred to exist in these planets even now.

## 2.2 Finding Critical value of Surface Heat flux for sustaining volcanism in rocky planetary bodies

### 2.2.1 *Internal heat parameters*

In this section we will describe the procedure and results of our inference of current internal heat parameters ( total internal heat in TW , surface heat flux in W/m$^2$ and total radiogenic heat in W/m$^2$ ) of planetary objects in the inner solar system : Earth,Moon,Mars,Venus and Mercury.

### *2.2.1 Earth*

The total internal heat generated in earth is estimated as 47±2 TW based upon 38347 heat flow measurements distributed over the earth's crust(Davies et al.,2010).The Radiogenic heating in earth is estimated from geoneutrino experiments(Araki et al.,2005). The decay of radiogenic isotopes *viz.*Uranium, Thorium and Potassium, (U 238,Th 232 and K40)in the planet's interior provides a continuing heat source which helps to cool the Earth. Radiogenic decay process inside the earth can be inferred from the flux of geoneutrinos which are electrically neutral particles that are can pass through the Earth virtually not affecting the surroundings. From combined measurements of the geoneutrino flux from the Kamioka Liquid-Scintillator Antineutrino Detector, Japan, with and measurements from the Borexino detector, Italy it is found that decay of U 238 and Th 232 together contributes about 20.0 TW to Earth's internal heat flux. The neutrinos emitted from the decay of K40 are inferred to contribute around 4 TW (Araki et al.,2005; Gando et al.,2011; Ludhova and Zavatarelli, 2013). From the best estimates of internal heat flow in earth 47 TW and the results of total radiogenic heat production from geoneutrino measurements 24 TW Urey ratio in Earth is inferred of the order of 0.5

### *2.1.2 Venus*:

The total surface heat flow is inferred to be of the order of 80TW (https://www.univie.ac.at/physikwiki/images/e/e7/Shf_venus.pdf) and the radiogenic heat of Venus is assumed to be 90 %of that of Earth (Ghail.,2015).

### 2.1.3 Mars

The present day surface heat flux estimation of mars is based on some of the models since it is difficult to make direct measurements on this planetary body. Some of the models were based on the thermal state of the lithosphere and its mechanical strength and some others based on the theoretical modeling of the internal heat evolution(Ruiz 2014) .Here in our paper we used the average heat production in Mars which is of the order of 19 mW/m$^2$ .This value obtained from the theoretical modeling of radiogenic heat production of crust and mantle and also on the scaling of heat flow variations which arises from the crustal variations and topography variations and on the heat flow based on relative elastic thickness of lithosphere beneath the North polar regions(Laura et al.,2017).The heat flow varies in between 14 mW/m$^2$ and 25mW/m$^2$, with an average of 19 mW/m$^2$.

### 2.1.4 Mercury

The composition of mercury might be best represented by a chondritic composition with a greater inventory of volatiles. The average value of initial heat production per kg from Cl chondrite model and from EH chondrite model is $Q_0=3.6*10^{-11}$ W/kg(Patrick et al.,2011; Nathalie et al.,2013).From this we calculate the value of heat production from the radioactive decay law $Q =Q_0 \exp(-\lambda t)$.The obtained value of Q by putting the value of λ as $1.5*10^{-17}$ ,we get the value of surface heat production in TW as 1.423 TW.

### 2.1.5 Moon:

A distribution map of lunar surface heat flow is derived from calibrated Kaguya Gamma Ray spectrometer data that is the abundance of K,U and Th(Yamashita et al.,2010).It can be seen that lunar surface heat flow are as low as 10.6 mW/m$^2$ in Polar Regions while they are up to 66.1mW/m$^2$ near the equator. The average surface heat flow is inferred as 18.3mW/m$^2$ and the Urey ratio is assumed to be 0.5(Dan Zhang et al.,2014).By using the Urey ratio the radiogenic heat is obtained to be as 9.15 mW/m$^2$ which is converted to TW to obtain 0.34 TW.

In Table 1 we have given the values of inferred current internal hear parameters of the different planetary objects in the inner solar system

## 2.2 The time evolution (thermal history) of internal heat parameters and volcanic activity of inner solar system planetary objects

Assuming a value of Urey ratio for inner solar system planetary objects and also radioactive decay laws it is possible to infer the time evolution of internal heat parameters of these objects over geological time scales.

The time evolution of total radiogenic heat ($R_H$) of a planetary object obeys the relation:

$$R_H(t) = R_H(0) \exp(-\lambda t) \qquad (1)$$

Here $R_H(t)$ is the total radiogenic heat in the planetary object at time t and $R_H(0)$ is the radiogenic heat at the time of formation of the planet. Here the time is reckoned in Ga (Giga years).

The current age of all planetary objects in the inner solar system is assumed to be the age of the Earth.

If $R_H(4.5)$ is equal to the value of the current inferred value RH of the planetary object given in Table 2

then can calculate $R_H(0)$ of the planetary object using (1) from:

$$R_H(0) = R_H(4.5) / \exp(-\lambda t) \qquad (2)$$

Here t=4.5 Gyrs.

Again using (1) we can calculate RH (t) of any t.

Assuming a value of 0.5 for the Urey ratio of the planetary object we can find the total internal heat for any time t from

$$I_H(t) = 2 R_H(t) \qquad (3)$$

If R is the radius of the given planetary object then the surface heat flux (S) in that object can be calculated from:

$$S(t) = I_H(t) / 4\pi R^2 \quad (4)$$

Using the above relations (1) to (4) we have calculated S (t) for different planets since the time of its formation ( t=0) and up to 12 Gyrs of age ( t=12) in steps of 0.5 Ga. The results are shown in Fig 1 to Fig 5 for different inner planetary objects which are essentially plots of time evolution of internal heat in these objects. The best fit exponential curves are also shown in these Figures.

### *2.3. Value of surface heat flux at the time of cessation of major volcanic activity for rocky planetary objects*

The details of volcanic activity in the past and present of different planetary objects in the inner solar system are given in Table 1. In the case Venus the major volcanic activity ceased around 2.5 Mya ,the surface heat flux value corresponding to that period is 0.091 W/m$^2$.For Mars the major volcanic activity is inferred to have ceased in the Hesperian Epoch ,around 3.5Gya ,and during this period we have inferred S values for this planets as 0.094W/m$^2$.The results are given in Table 3.In the case of Mercury the major volcanic activity ceased around 3.5Gya,the surface heat flux value corresponding to this period is 0.094W/m$^2$.In the case of Moon the major volcanic activity ceased around 3.3Gya ,the surface heat flux value corresponding to this period is 0.085W/m$^2$.The mean value corresponding to these values is around 0.09225 W/m$^2$ . This value is close to the current S value of Earth so that it may be the critical internal heal flux of any rocky planetary object to sustain major volcanism in the same.

### *2.4 Inferring the period of possible cessation of volcanic activity in Earth*

The evolution of volcanic activity in Earth could have influenced the origin and evolution of life forms in this planet. Advanced life forms (plant or animal life) are inferred to have appeared in Earth only 500 Myrs (Hessen 2008). In Fig 6 we have plotted maximum inferred intensity of volcanic eruptions in four geological periods expressed in VEI scale. The recent volcanic eruption data is given in Table 4 and the details of five major mass with associated volcanic eruption details if any in Table 5.The maximum VEI inferred in Earth 27.8 Myr is found to 9.2 and during the current Holocene period to be 8. The volcanic eruption around 65

Myr which is associated with Deccan KT extinction (mass extinction of Dinosaurs) must be at least one order of magnitude higher than that for period 3. CAMP volcanic eruption is still larger spatial and temporal dimensions and is given a maximum value 11. The mass extinctions during periods further back in time such as during Permian-Triassic ( 251 Ma), Late Devonian period (364 Ma) and Ordovician periods (439 Ma) are possibly associated with volcanic eruptions in Earth with maximum VEI ≥11 (Blackburn T J, Olsen P E et al., 2013) .Thus we could infer possibly steady decline in the magnitude of volcanic eruptions in Earth during the past 500 Ma in general and past 200 Ma in particular when advanced life forms evolved in Earth. The notable absence of major mass extinctions after the KT extinction (past 65 Ma) helped the development of human and intelligent life in Earth. These results also support volcanism cessation hypothesis in near geological future as inferred by us and explained in this paper from the relevant observations of other rocky planetary objects in our solar system.

From 500 Myrs in the past to at present the S value of Earth is inferred to have decreased from 0.116 to 0.093 W/m$^2$.Assuming a linear change (Fig 1) this implies a change of 4.6 X 10$^{-5}$ W/m$^2$ per Ma. During the time of extinction of Dinosaurs (assumed to be due to a major volcanic eruption) 65 Myrs (Graeme et al.,2008; Keller et al.,2009) the value S can be inferred to be 0.096 W/m$^2$ This period may be the starting of decline  of major volcanic activity in Earth. Venus, the twin planet of Earth in mass shows no volcanic activity for the past 2.5 Ma .When the S value of earth decreases to 0.091 W/m$^2$ (the inferred S value of Venus at the time cessation in Table 2) the major volcanism in our planet is most likely to disappear. This happens in a time of 65.2 Ma years from now.  The mean value of S at the time of cessation inferred for rocky planetary object from Table   is found to be 0.0923 W/m$^2$. The geological age at which Earth attains this S value can be inferred to be 19 million years from now.

## 3 Conclusions

In this paper we have proposed the hypothesis that every rocky planet requires a critical value to sustain major volcanic activity in the same and when its surface heat flux reduces to a value below this critical value there is every possibility of cessation of volcanism in the planet. Earth is the only rocky planetary object in our Solar system to have sustained volcanism over the geological time scales since its formation. The current surface heat flux of Earth (0.093 W/m$^2$) is inferred to be the critical value of S required to sustain major volcanic activity in a rocky planet. This result is based on our analysis of existing information on the thermal and volcanic activity history of rocky planetary objects in the inner solar system. Our value is at least two times larger than similar critical value for volcanism suggested by Williams et al.1997.

We have found some evidence for possible decline of the magnitude of volcanic eruptions in Earth since the KT extinction event associated with Deccan volcanic eruption in India 65 million years ago. Apart from intensity reduction in spatial or latitudinal extent of volcanism may another indicator for decline of volcanic activity in a rocky planet. In Table 6 we have given maximum latitudinal extent of volcanic eruptions observed during the geological past in different rocky planetary objects in our solar system. Maximum latitudinal extent of major volcanism in Earth has reduced from 84 degrees to 65 degrees from geographic equator. In a similar manner in Venus also we can find a significant reduction in the spatial extent of volcanism. It is inferred by us that major volcanic activity in Earth may cease between 19-65 million years from now.

The geology and atmospheric conditions in Mars changed drastically when major volcanism ceased in the red planet approximately at 3.5Myrs. The large volumes of surface water in Mars once existing (Halevy et al.2014;Linda 2010) was a byproduct of intense volcanic activity ( it had largest shield volcanoes in the solar system history) and at present it has only traces of near surface water due to minor volcanic activity. In a similar way we can expect drastic changes in Earth after the cessation of volcanic/tectonic activity in its interior, atmosphere and biosphere.

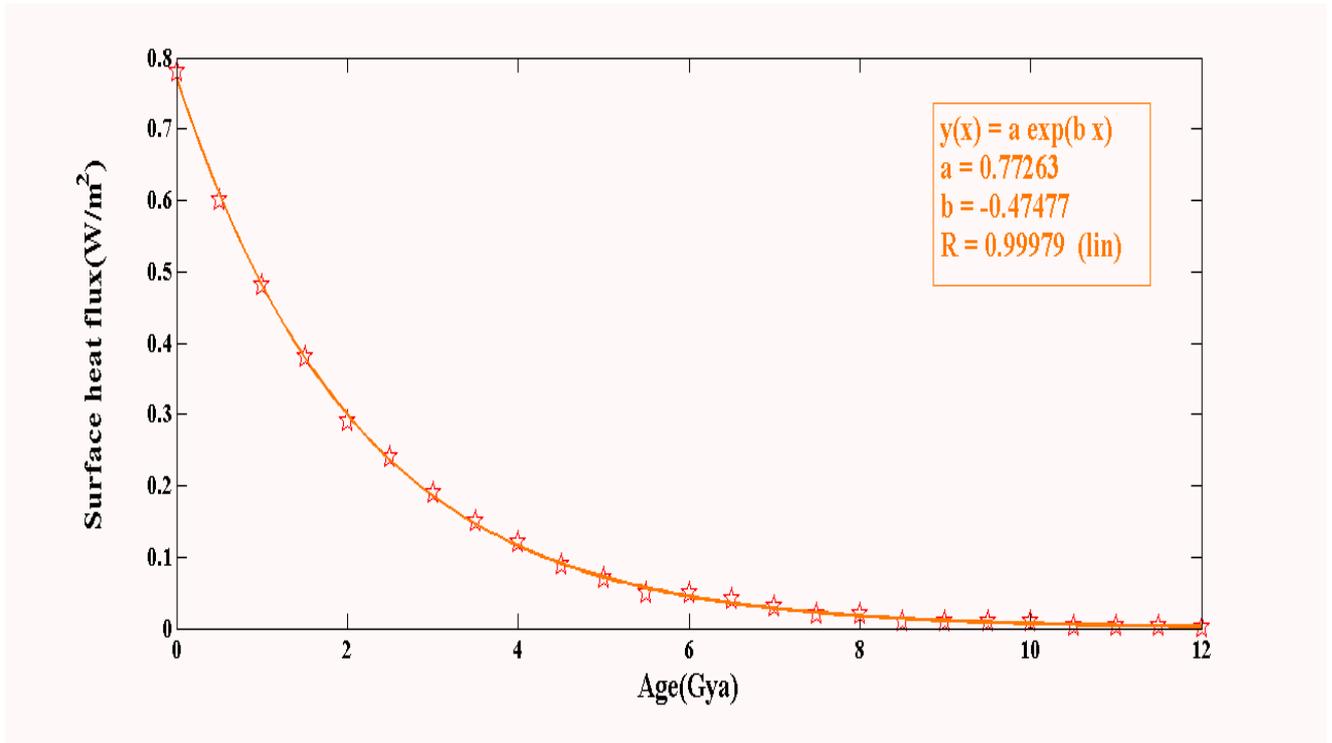

**Figure 1: Volcanic history plot for Earth**

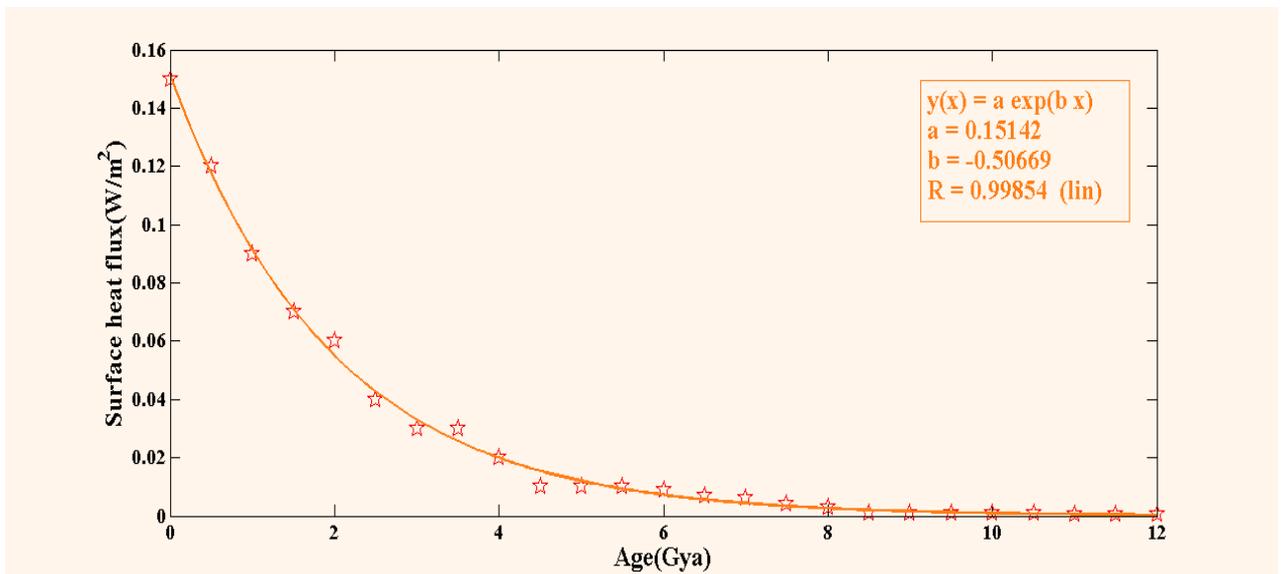

**Figure 2: Volcanic history plot for Venus**

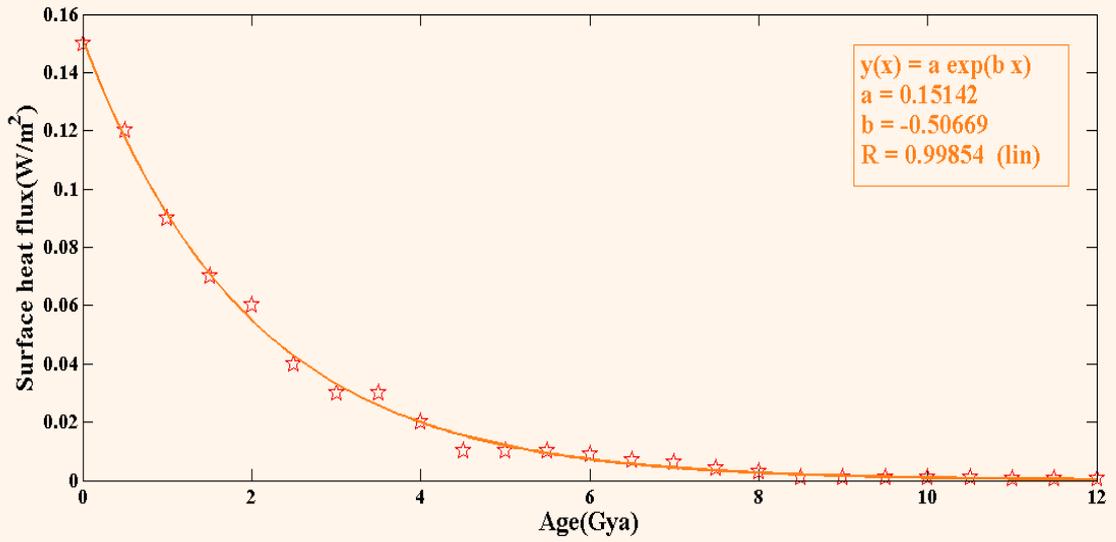

**Figure 2: Volcanic history plot for Mars**

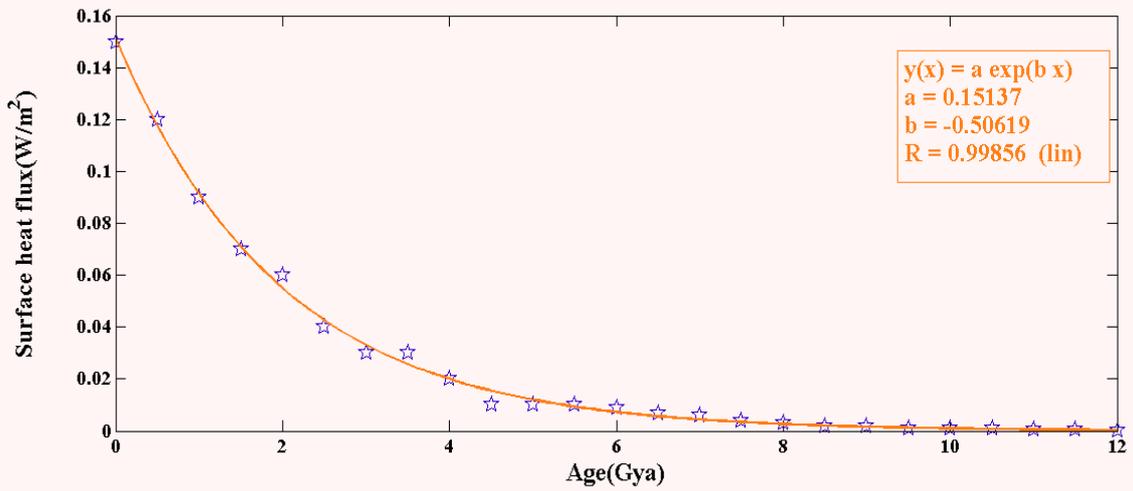

**Figure 3: Volcanic history plot for Mercury**

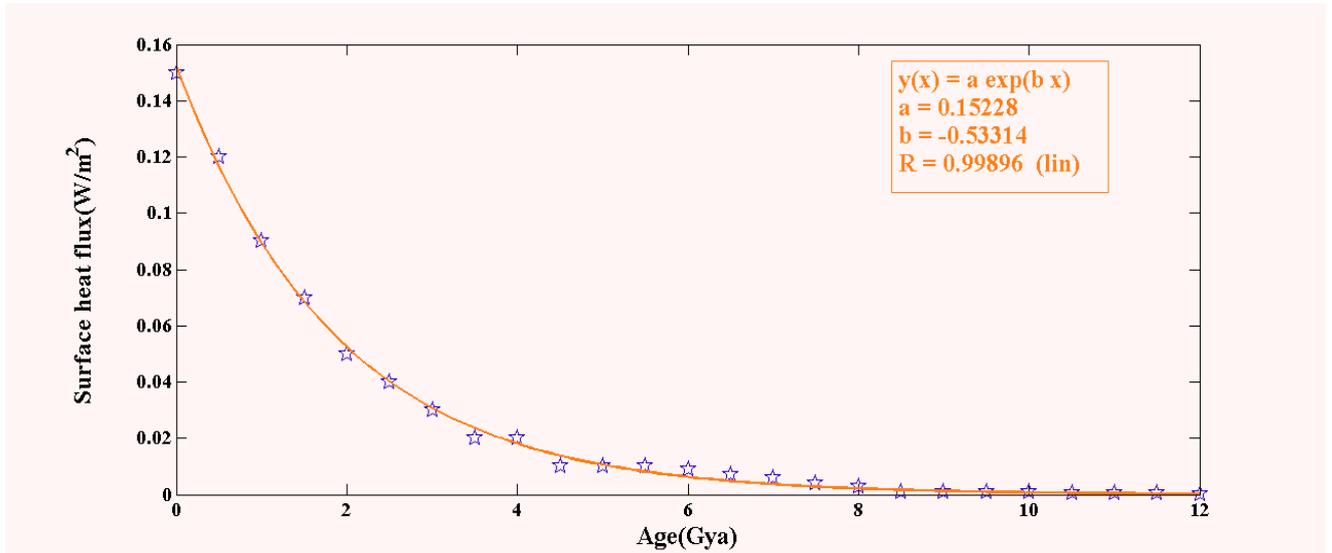

**Figure 4:Volcanic history plot for Moon**

**Table 1 Volcanic activity related data available for rocky planetary objects in the inner solar system**

| Planet | Period of past major volcanism | Present minor volcanism if any | Active major volcanoes in the present or past |
|--------|-------------------------------|--------------------------------|-----------------------------------------------|
| Earth | Continuing | Minor volcanism | 1450 current volcanoes |
| Venus | 2.5 Mya | Detected in Regions whose latitude varies from $24.3^0$S to $46.5^0$S. | Around 1500 active major volcanoes including in the geological past. |
| Mars | 3.7 to 3.5Gya | Recently eight lava flows detected in three shield clusters (South East of Olympus Mons, Ceranius Fossae, South East of Pavonis Mons). The latitude of identified lava flow extended from $23.65^0$N to $0.6^0$S. | 24 major volcanoes. Olympus Mons is the largest volcano in the solar system. |
| Mercury | Around 3.5Gya | | |
| Moon | 3.3Gya | | Number of shield volcanoes in the geological past is identified to be 8. |

**Table 2: Internal Heat parameters of rocky planets**

| Planetary body | Total surface heat $I_H$(TW) | Surface Heat Flux(W/m$^2$) | Radiogenic Heat RH(TW) |
|---|---|---|---|
| Earth | 47 | 0.093 | 23.50 |
| Venus | 42.3 | 0.092 | 21.15 |
| Mercury | 1.42 | 0.018 | 0.71 |
| Mars | 2.74 | 0.019 | 1.37 |
| Moon | 0.69 | 0.018 | 0.35 |

**Table 3: Period of Cessation of Volcanic activity in planetary body and the corresponding surface heat flux from the graph**

| Planetary body | Period of Cessation of major volcanic activity from current epoch(Gya) | Surface heat flux inferred during this period(W/m$^2$) |
|---|---|---|
| Venus | 2.5 Million year ago | 0.091 |
| Mars | 3.5 | 0.099 |
| Mercury | 3.5 | 0.094 |
| Moon | 3.3 | 0.085 |
| | | Average Value of heat flux=0.09225W/m$^2$ |

**Table 4: Frequency of occurrence of extreme volcanic events in different geological sub periods during the past 500 Mya with different VEI values**

| Period | Epoch | VEI=7 | VEI=8 | VEI 8.1 -8.9 | VEI>9 |
|---|---|---|---|---|---|
| Quaternary | Holocene(11,700 years to present) | 3 | 1 | 0 | 0 |
| | Pliestocene(2.588Mya to 11,700 years) | | | 7 | |
| Neogene | Pliocene(5.332 to 2.58Mya) | | | 4 | |
| | Miocene(23.03 to 5.332 Mya) | | | 8 | |
| Paleogene | Oligocene(33.9 to 23.03Mya) | | | 17 | 2 |
| | Eocene(55.8 to 33.9Mya) | | | 2 | |
| | Paleocene(66.5 to 55.8Mya) | | | | |
| Cretaceous | Upper Cretaceous(99.6 to 65.5Mya) | | | 1 | |
| | Lower Cretaceous (145.5 to 99.6Mya) | | | 1 | |
| Paleozoic | Upper Ordovician (460.9Mya to 443.7Mya) | | | 3 | |

**Table 5: Major mass extinction events during the past 500 Myr and associated volcanic eruption details if any**

| Sl No. | Major Mass Extinction Events | Period | Area of coverage of associated volcanic eruption if any(Mkm$^2$) |
|---|---|---|---|
| 1 | Cretaceous –Paleogene extinction event | 65 million years ago | >1 |
| 2 | Triassic –Jurassic Extinction Event | 199 Million to 214 Million years ago | >7 |
| 3 | Premian –Triassic Extinction event | 251 Million years ago | >4 |
| 4 | Late Devonian Extinction | 364 million years ago | |
| 5 | Orodovician –Silurian Extinction events | 439 Million years ago | |

**Table 6 Spatial extend of volcanism in the inner solar system**

| Planet | Maximum latitudinal extend of volcano |
|--------|---------------------------------------|
| Earth  | $84^0$N to $87^0$S |
| Venus  | $78.5^0$N to $60^0$S |
| Mars   | $48^0$N to $36^0$S |
| Moon   | $161^0$N to $8^0$N |